\documentclass[%
 reprint,
superscriptaddress,
amsmath,amssymb,
aps,
]{revtex4-2}

\usepackage{amssymb}
\usepackage{amsmath} 
\usepackage{float}
\usepackage{graphicx}
\usepackage{dcolumn}
\usepackage{bm}
\usepackage{subcaption}
\usepackage{caption}
\usepackage{textcomp} 
\usepackage{xcolor}
\usepackage{gensymb}

\usepackage{hyperref}       
\usepackage{xcolor}         

\hypersetup{
    colorlinks=true,        
    linkcolor=blue,         
    citecolor=blue,          
    urlcolor=magenta        
}

\begin{document}

\preprint{APS/123-QED}
\title{Accelerating Bayesian Sampling for Massive Black Hole Binaries \\
with Prior Constraints from Conditional Variational Autoencoder}

\author{Hui Sun}
\email{sunhui22@mails.ucas.ac.cn}
\affiliation{University of Chinese Academy of Sciences (UCAS), Beijing 100049, China
}%
\affiliation{International Centre for Theoretical Physics Asia-Pacific, UCAS, Beijing 100190, China
}%
\affiliation{Taiji Laboratory for Gravitational Wave Universe (Beijing/Hangzhou), UCAS, Beijing 100190, China
}
\author{He Wang}
\email{hewang@ucas.ac.cn}
\affiliation{University of Chinese Academy of Sciences (UCAS), Beijing 100049, China
}%
\affiliation{International Centre for Theoretical Physics Asia-Pacific, UCAS, Beijing 100190, China
}%
\affiliation{Taiji Laboratory for Gravitational Wave Universe (Beijing/Hangzhou), UCAS, Beijing 100190, China
}
\author{Jibo He} 
\email{jibo.he@ucas.ac.cn}
\affiliation{University of Chinese Academy of Sciences (UCAS), Beijing 100049, China
}%
\affiliation{International Centre for Theoretical Physics Asia-Pacific, UCAS, Beijing 100190, China
}%
\affiliation{Taiji Laboratory for Gravitational Wave Universe (Beijing/Hangzhou), UCAS, Beijing 100190, China
}
\affiliation{Hangzhou Institute for Advanced Study, UCAS, Hangzhou 310024, China
}

\date{\today}

\begin{abstract}
A Conditional Variational Autoencoder (CVAE) model is employed for parameter inference on gravitational waves (GW) signals of massive black hole binaries, considering joint observations with a network of three space-based GW detectors. Our experiments show that the trained CVAE model can estimate the posterior distribution of source parameters in approximately one second, while the standard Bayesian sampling method, utilizing parallel computation across 16 CPU cores, takes an average of 20 hours for a GW signal instance. However, the sampling distributions from CVAE exhibit lighter tails, appearing broader when compared to the standard Bayesian sampling results. By using CVAE results to constrain the prior range for Bayesian sampling, the sampling time 
is reduced by a factor of $\sim$6 while maintaining the similar precision of the Bayesian results.
\end{abstract}

\maketitle


\section{introduction}

The first detection of gravitational waves (GW)~\cite{abbott2016observation} marked the beginning of a new era in astronomical and cosmological observations. Ground-based GW detectors have already identified numerous mergers of compact objects \cite{abbott2019gwtc, abbott2021gwtc, abbott2023gwtc}. Space-based GW detectors, 
such as Taiji~\cite{hu2017taiji}, LISA~\cite{amaro2017laser}, and TianQin~\cite{luo2016tianqin}, 
are capable of detecting GWs at significantly lower frequencies compared to the ground-based detectors. Jointly considering signals from multiple space-based GW detectors can enhance the sensitivity and reliability of GW detection~\cite{gong2021concepts, zhang2021sky, hu2021joint}. 

The massive black hole binary (MBHB) system with component masses ranging from $10^5\,\mathrm{M}_{\odot}$ to $10^7\,\mathrm{M}_{\odot}$ is~one of the primary GW sources for the space-based detectors. The detection~\cite{harry2008hierarchical} and parameter estimation~\cite{marsat2021exploring,hu2021joint} of MBHB signals are fundamental topics of this field. Bayesian inference, as a fundamental methodology in gravitational wave (GW) analysis, relies on Bayes\text{'} theorem to estimate the posterior probability of source parameters. The Bayes\text{'} theorem is expressed as

\begin{equation}
p(\theta|d)=\frac{\mathcal{L}(d|\theta)\pi(\theta)}{\mathcal{Z}},
\label{eq:my}
\end{equation} 
where $\theta$ denotes the GW parameters under consideration, and $d$ represents the strain data as a superposition of detector noise and the detector's response to GW signal. The term $\mathcal{Z} = \int \mathcal{L}(d|\theta) \pi(\theta) d\theta$ is the evidence, treated as a constant and often negligible. The $\pi(\theta)$ is the prior distribution, and $\mathcal{L}(d|\theta)$ is the likelihood.

Sampling methods based on Bayesian inference, such as Markov Chain Monte Carlo (MCMC)~\cite{hastings1970monte,christensen2001using,raymond2010effects} and Nested Sampling (NS)~\cite{skilling2006nested}, are well implemented with high sampling precision, but they can be computationally demanding. Recent studies have exploited the use of deep learning model for rapid posterior estimation of source parameters of stellar-mass black hole binaries in ground-based GW observations, such as Conditional Variational Autoencoder (CVAE)~\cite{gabbard2022bayesian} and normalizing flows~\cite{green2021complete}. However, the current application of normalizing flows for parameter estimation of MBHB systems tends to yield broader posterior distributions compared to the Bayesian inference method, with the discrepancy being approximately an order of magnitude~\cite{Du:2023plr}.

For stellar-mass black hole binary sources, the posterior distributions estimated by neural networks generally match the standard Bayesian sampling results in terms of spread, but can exhibit noticeable discrepancy in morphology. To address this issue, a method known as Neural Importance Sampling has been proposed, which reweights the samples generated by deep learning model to correct inaccuracies in the estimated distributions~\cite{dax2023neural}. However, overly broad proposals can still result in low sample efficiency~\cite{dax2023neural}, limiting the applicability of this method.

Inspired by recent work on hierarchical parameter space reduction for extreme-mass-ratio inspirals (EMRIs) using both physical and phenomenological waveforms~\cite{ye2024identification}, we adopt a similar idea for MBHB signals. Specifically, we extract boundary information from the CVAE-generated samples while ignoring the detailed shape information of the distribution. This boundary information is then applied as a narrowed prior in Bayesian sampling, accelerating the standard Bayesian sampling process. 

Other methods utilizing GW properties have also been proposed to accelerate the Bayesian sampling process of GW parameter estimation, such as Relative Binning~\cite{zackay2018relative,leslie2021mode}, Multibanding~\cite{vinciguerra2017accelerating}, and Reduced Order Quadrature~\cite{canizares2015accelerated}. These methods achieve acceleration by modifying the likelihood rather than the prior.

This paper is organized as follows. The methodology is provided in Sec.~\ref{sec:meth}, with a brief introduction to the response mechanism of the space-based detectors network in Sec.~\ref{sec:space}, as well as the fundamental principle and sampling method of standard Bayesian inference in Sec.~\ref{sec:ns} and CVAE in Sec.~\ref{sec:cvae}. The experiments are presented in Sec.~\ref{sec:train}, including the training process and testing result of CVAE in Sec.~\ref{sec:cvaetrain} and comparison of Nested Sampling results before and after applying the narrowed prior in Sec.~\ref{sec:nsresults}. In Sec.~\ref{sec:conclu}, we provide the conclusion of the paper.

\section{methodology}
\label{sec:meth}
\subsection{Space-Based Interferometers Network}
\label{sec:space}

We consider a network of three space-based GW detectors~\cite{cai2024networks}: Taiji, TianQin, and LISA. The centers of these detectors all follow the Earth-like orbit, each separated by $20\degree$ relative to the orbital center, with a period of one year. Each space-based GW detector have two independent data channels, denoted as A and E channels. Therefore, we will analyze a total of 6 independent channels in the detectors network.

The response of each channel to the plus component $h_+$ and the cross component $h_\times$ of the GW is described as
\begin{equation}
s(t)=F_+h_+(t)+F_{\times}h_{\times}(t), 
\end{equation} 
where $F_+$ and $F_\times$ are the polarization response functions of the channel:
\begin{equation}
F_+=D^{ij}e_{ij}^+, \;F_\times=D^{ij}e_{ij}^{\times}. 
\end{equation} 
$D^{ij}$ is the detector tensor of the channel and $e_{ij}$ is the GW polarization tensor. The detector tensor $D^{ij}$ is related to the directions of the three arms of the detector. The $D^{ij}$ values of the A and E channels are calculated as~\cite{marsat2021exploring} 
\begin{align}
D_A^{i j} & =\frac{1}{6}\left(u_1^i u_1^j-2 u_2^i u_2^j+u_3^i u_3^j\right),  \\
D_E^{i j} & =\frac{\sqrt{3}}{6}\left(u_1^i u_1^j-u_3^i u_3^j\right), 
\end{align}
where $u_1^i, u_2^i, u_3^i$ represent the $i$-th components of the arm direction vectors for the three arms. The GW polarization tensor $e_{ij}$ is relevant to the ecliptic latitude and ecliptic longitude of the GW source in the ecliptic coordinate system, as well as the polarization angle of the GW. The detailed expression of the polarization response functions $F_+$ and $F_\times$ can be found in Ref.~\cite{liang2019frequency}.

\subsection{Nested Sampling} 
\label{sec:ns}

In the context of Bayesian inference, as expressed in Eq.~\eqref{eq:my}, the likelihood function $\mathcal{L}(d|\theta)$ is defined as
\begin{equation}
\mathcal{L}(d|\theta)=\mathrm{exp}\left[-\frac{1}{2}\left\langle d- s(\theta)|d- s(\theta)\right\rangle\right]. 
\end{equation} 
Here, $s(\theta)$ represents the detector's response to the GW signal characterized by parameters $\theta$. The inner product is defined as 
\begin{equation}
\left\langle {a}|{b}\right\rangle=4\cdot\mathfrak{R}\left[\int_0^{\infty} \frac{\tilde{a}(f)\tilde{b}^*(f)}{S_n(f)}df\right], 
\end{equation} 
where $\tilde{a}(f)$ and $\tilde{b}(f)$ denote the Fourier transforms of $a(t)$ and $b(t)$, $\mathfrak{R}$ represents the extraction of the real part from a complex number, and $\tilde{b}^*(f)$ denotes the complex conjugate of $\tilde{b}(f)$. $S_n(f)$ is the one-sided power spectral density, following the model in Ref.~\cite{liu2020exploring} for Taiji and TianQin, and Ref.~\cite{niu2020constraining} for LISA.

Nested Sampling is a classic sampling method for Bayesian inference, first proposed by Skilling~\cite{skilling2006nested}. This method randomly generates sampling points of $\theta$ according to the prior $\pi(\theta)$ and iteratively selects $\theta_i$ in the $i$-th iteration corresponding to a smaller prior volume $X$, which is defined as
\begin{equation}
X(\mathcal{L})\equiv\int_{\mathcal{L}(\theta)>\mathcal{L}} \pi(\theta)d\theta. 
\label{eq:volume}
\end{equation}
The definition of $X$ in Eq.~\eqref{eq:volume} implies that a smaller $X_i$ corresponds to a higher likelihood $\mathcal{L}_i$. The importance weighting $\hat{P}_i$ of $\theta_i$ is then calculated based on the prior volume $X_i$ estimated through statistical properties and the likelihood $\mathcal{L}_i$ computed from the likelihood function $\mathcal{L}(d|\theta)$, expressed as
\begin{equation}
\hat{P}_i = \frac{\mathcal{L}_i(\hat{X}_{i-1}-\hat{X}_i)}{\sum_i \mathcal{L}_i(\hat{X}_{i-1}-\hat{X}_i)}. 
\end{equation}
Each sampling point $\theta_i$ is weighted by its importance weighting $\hat{P}_i$ to construct the target posterior distribution.

\begin{figure}
\includegraphics[width=8.5cm]{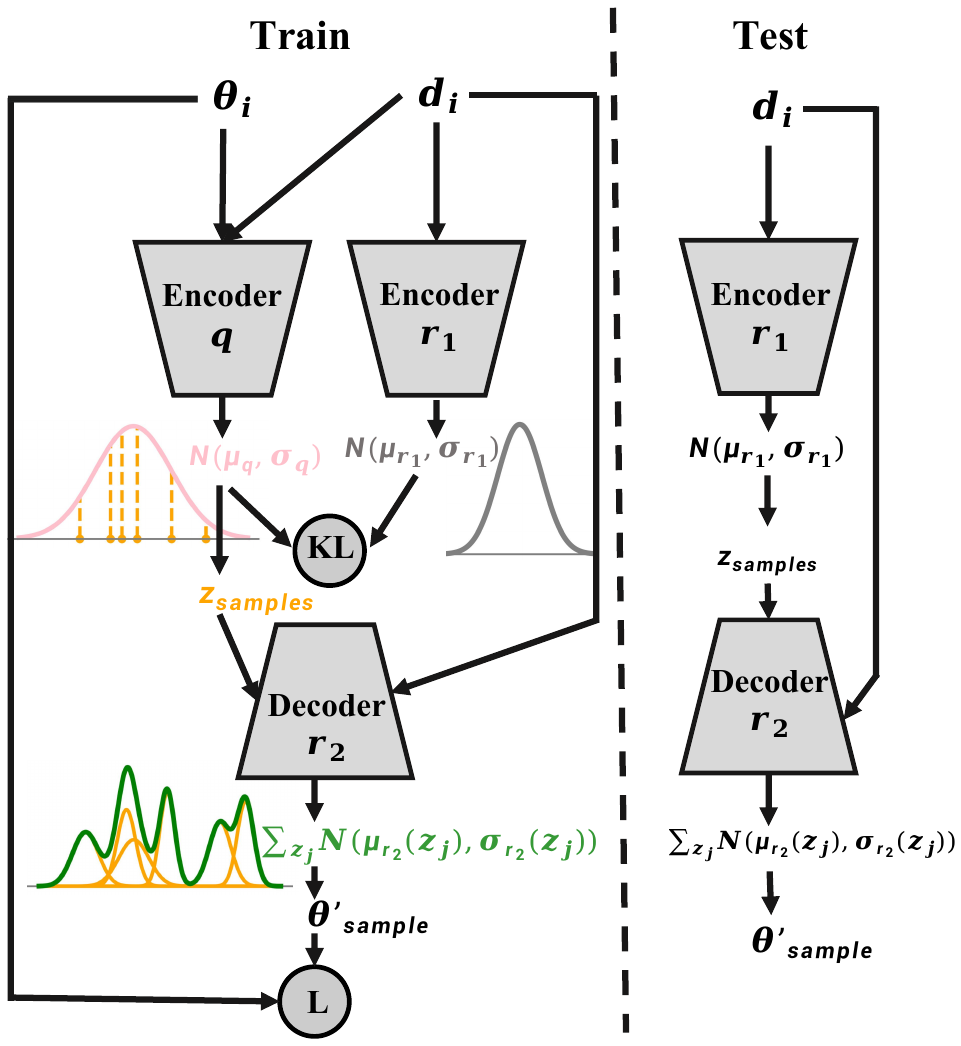}
\captionsetup{justification=raggedright,singlelinecheck=false}
\caption{\label{fig:network} The encoder-decoder architecture of CVAE designed for posterior inference. During training (left), both GW parameters $\theta_i$ and time-domain signal $d_i$ are the inputs to the network. The encoders $q$ and $r_1$ generate the normal distribution $N(\mu_q,\sigma_q)$ and $N(\mu_{r_1},\sigma_{r_1})$ in the latent space, and $z_{\rm{samples}}$ are sampled from $N(\mu_q,\sigma_q)$. These samples are then the inputs to the decoder $r_2$ along with $d_i$, producing a complex distribution approximating the true Bayesian posterior distribution $p(\theta|d_i)$. The KL divergence $D_{\rm KL}$ is used to measure the discrepancy between the distributions $N(\mu_q,\sigma_q)$ and $N(\mu_{r_1},\sigma_{r_1})$. The training loss is computed as the sum of the reconstruction error from the sampled $\theta^{\prime}$ to the ground truth $\theta_i$, and the KL divergence. During testing (right), by sequentially inputting $d_i$ into encoder $r_1$ and decoder $r_2$, we can obtain samples $\theta^{\prime}$ to estimate the posterior of $d_i$. }
\end{figure}

The Nested Sampling procedure is as follows. Initially, $n_{\mathrm{live}}$ samples are drawn from the prior $\pi(\theta)$ to act as live points in the first iteration. Subsequently, the sample with the lowest likelihood among the live points in the $i$-th iteration is removed as a dead point $\theta_i$, with the prior volume of the dead point statistically estimated as $\hat{X}_i = \mathrm{exp}(-i/n_{\mathrm{live}})$~\cite{skilling2006nested}. This choice is based on the fact that the ratio of prior volumes between iterations, $t_i = X_i / X_{i-1}$, follows a Beta distribution, specifically denoted as $t_i \sim \beta(n_{\mathrm{live}}, 1)$. A new live point is then sampled from a constrained prior that encloses the remaining live points, with the condition that its likelihood must exceed that of the most recently removed live point. This iterative process continues until the sampling precision meets the stopping criterion. The evidence $\mathcal{Z}$ is then calculated as $\sum_i \mathcal{L}_i(\hat{X}_{i-1}-\hat{X}_i)$. A typical stopping criterion is when the remaining evidence among live points, estimated as the product of the current maximum likelihood among the live points and prior volume of the latest removed live point, falls below a small fraction of the accumulated evidence among dead points. 

Several samplers have been developed based on the Nested Sampling framework, such as \texttt{PyMultiNest}~\cite{buchner2014x, feroz2008multimodal, feroz2009multinest}, \texttt{dynesty}~\cite{speagle2020dynesty}, \texttt{nestle}~\cite{barbary_nestle}, and \texttt{PolyChord}~\cite{handley2015polychord}. In this work, we use the \texttt{PyMultiNest} sampler, which is well parallelized, allowing us to perform multi-core CPU sampling. Specifically, we employ 16 CPU cores for parallel sampling.

\subsection{CVAE model} 
\label{sec:cvae}

In this paper, we utilize the Conditional Variational Autoencoder (CVAE) model~\cite{sohn2015learning}, comprising two encoders denoted as $r_1$ and $q$, and a decoder denoted as $r_{2}$, to approximate the target Bayesian posterior distribution. The architecture of the CVAE is illustrated in Fig.~\ref{fig:network}. During the training procedure depicted in Fig.~\ref{fig:network}, the strain data $d_i$ and the corresponding GW source parameters $\theta_i$ from the $i$-th data instance are fed into the encoder $q$. The encoder $q$ then generates the mean $\mu_q$ and standard deviation $\sigma_q$ which define a normal distribution $N(\mu_q,\sigma_q)$ for the latent variable $z$. Samples are drawn from the distribution $N(\mu_q,\sigma_q)$ to generate $z_{\rm{samples}}$. 

Each $z_{j}\in z_{\rm{samples}}$ is an input to the decoder $r_2$ along with $d_i$, producing a mean $\mu_{r_2}(z_j)$ and standard deviation $\sigma_{r_2}(z_j)$ to construct a normal distribution $N(\mu_{r_2}(z_j),\sigma_{r_2}(z_j))$ for $\theta^{\prime}$. By combining all normal distributions corresponding to $z_{\rm{samples}}$, a complex distribution can be constructed to approximate the true Bayesian posterior distribution $p(\theta|d_i)$. However, this procedure requires the input to include $\theta_i$, which is unavailable when testing on $d_i$ with unknown $\theta_i$. 

To address this issue, the model introduces the encoder $r_1$. By inputting only $d_i$ into the encoder $r_1$ to obtain the normal distribution $N(\mu_{r_1},\sigma_{r_1})$ and training the model to minimize the discrepancy between $N(\mu_{r_1},\sigma_{r_1})$ and $N(\mu_q,\sigma_q)$, the encoder $r_1$ can be utilized during testing to generate the distribution of the latent variable $z$ instead of the encoder $q$, as depicted in the testing procedure of Fig.~\ref{fig:network}.

The discrepancy between the distribution $r_{\rho_1}(z|d_i)\equiv N(\mu_{r_1},\sigma_{r_1})$ and $q_{\phi}(z|\theta_i,d_i)\equiv N(\mu_q,\sigma_q)$ can be assessed using the Kullback-Leibler (KL) divergence $D_{\rm KL}$, defined as
\begin{align}
D_{\rm KL}&\left(q_{\phi}(z|\theta_i,d_i)||r_{\rho_1}(z|d_i)\right) \\ \nonumber
&\equiv \int \mathrm{d}z\ q_{\phi}(z|\theta_i,d_i)\log\left[\frac{q_{\phi}(z|\theta_i,d_i)}{r_{\rho_1}(z|d_i)}\right].
\end{align}
The discrepancy between the true posterior $p(\theta|d_i)$ and the distribution $r_{\rho_2}(\theta^{\prime}|z,d_i)=\frac{1}{n}\sum_{j}^nN(\mu_{r_2}(z_j),\sigma_{r_2}(z_j))$ estimated by the model can be quantified using the reconstruction loss $-E_{q_{\phi}(z|\theta_i,d_i)}[\mathrm{log}r_{\rho_2}(\theta_i|z,d_i)]$. Here, the variable $n$ denotes the total number of $z_{\mathrm{samples}}$.

For the CVAE model, it can be demonstrated that the sum of the KL divergence loss and the reconstruction loss serves as an upper bound on the true loss of the model. A detailed proof of this statement can be found in Ref.~\cite{gabbard2022bayesian,tonolini2020variational}. When training the model, we sum these two losses to obtain the overall loss of the model.

\begin{figure*}[ht]
\includegraphics[width=17cm]{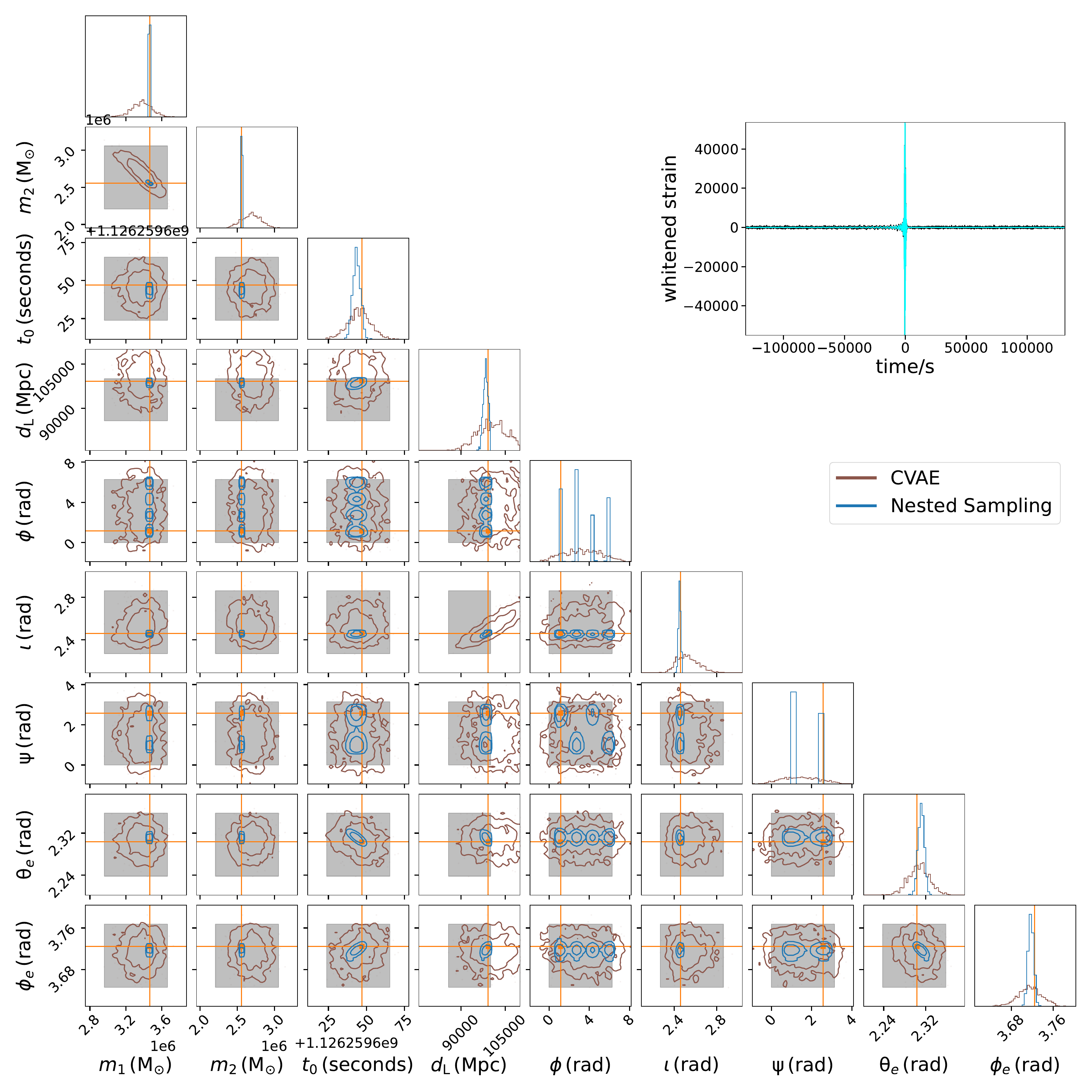}
\captionsetup{justification=raggedright,singlelinecheck=false}
\caption{\label{fig:corner} Distributions of source parameters obtained using the CVAE method (brown) and standard Bayesian sampling method (blue). The result contours of each method correspond to the $68\%$ and $95\%$ confidence levels. The histograms on the diagonal represent the posterior probability distribution for each parameter. The orange solid lines on the contour plot represent the true values of the parameters. We highlight in gray the constrained prior range, obtained by applying the CVAE-derived constraints within the initial full prior range. In the upper right, the whitened Taiji A-channel strains for this GW instance are shown, with the noiseless signal in light blue and the noisy signal in black. }
\end{figure*}

\section{TRAINING AND PRIOR NARROWING}
\label{sec:train}
\subsection{Data Generation and Training Model}
\label{sec:cvaetrain}

The training set comprises $6\times 10^5$ instances of GW data, each containing strain data from 6 channels across 3 detectors as features, and source parameters as labels. Strain data from each of the 6 channels is a time series with a data length of 16384, sampled at a frequency of 1/16 Hz, corresponding to a duration of approximately 3 days. The plus component $h_+$ and the cross component $h_\times$ are generated using the (2, 2) dominant mode of the GW under the $\texttt{IMRPhenomPv3}$~\cite{khan2019phenomenological} waveform approximation model. Each strain data is obtained by adding a response data from a MBHB GW waveform signal generated by \texttt{PyCBC}~\cite{biwer2019pycbc} to a random noise generated from the power spectral density $S_n(f)$ of each detector. The resulting strain data is further whitened by dividing it by the square root of the power spectral density $\sqrt{S_n(f)}$ to produce a dataset suitable for training. The validation set consists of $1.2 \times 10^4$ instances, and the test set consists of $50$ instances, both constructed in the same way as the training set. Distinct random seeds are employed for the training, validation, and test sets to ensure they are mutually exclusive. 

The GW source parameters $\theta$ of MBHB consist of 9 parameters, including the component masses $m_1$ and $m_2$, the GPS time $t_0$ at the merger of the binary, the luminosity distance $d_L$ from the source to the detector, the phase $\phi$ at the merger, the inclination angle $\iota$, the polarization angle $\psi$, the ecliptic latitude $\theta_e$, and the ecliptic longitude $\phi_e$ of the source in the ecliptic coordinate system. The default priors of these 9 parameters, including their distribution types and ranges, are specified in Table~\ref{tab:table1}. 

\begin{table}
\caption{\label{tab:table1} Default MBHB priors settings. The distributions of $\mathrm{cos(}\iota\mathrm{)}$ and $\mathrm{cos(}\theta_e\mathrm{)}$ are assigned to be uniform to reflect isotropy in space. Other parameters are set to be uniform in their specific ranges. The reference GPS time, denoted as $t_r$, is set to 1126259643\,$\mathrm{s}$. }
\begin{ruledtabular}
\begin{tabular}{cccc}
Variable &  Description & Prior & Range\\ \hline
$m_1$ & mass 1 & Uniform & $(10^5\,\mathrm{M}_{\odot}, 10^7\,\mathrm{M}_{\odot})$  \\
$m_2$ & mass 2 & Uniform & $(10^5\,\mathrm{M}_{\odot}, 10^7\,\mathrm{M}_{\odot})$ \\
$t_0$ & coalescence time &  Uniform &$(t_r - 256\,\mathrm{s}, t_r+ 256\,\mathrm{s}) $ \\
$d_L$ & luminosity distance & Uniform & $(10^3\,\mathrm{Mpc}, 10^5\,\mathrm{Mpc}) $ \\
$\phi$ & coalescence phase &  Uniform & $[0, 2\pi] $ \\
$\iota$ & inclination angle & Sine & $[0, \pi] $\\
$\psi$ & polarization angle &  Uniform & $[0, \pi] $ \\
$\theta_e$ & ecliptic latitude &  Sine & $[0, \pi] $ \\
$\phi_e$ & ecliptic longitude &  Uniform & $[0, 2\pi] $
\end{tabular}
\end{ruledtabular}
\end{table}

During training, we adopt a strategy of decreasing the learning rate as iterations increase, based on the InverseTimeDecay learning rate schedule defined as $r = r_0 / (1 + i / 10^5)$. Here, $r_0$ denotes the initial learning rate, which is fixed at $1 \times 10^{-4}$, and $i$ represents the training iteration. We train the model until the total loss is no longer decreasing significantly. This process takes two days on an A800 GPU, utilizing 68~GB of GPU memory and completing $\sim2 \times 10^5$ iterations.

The parameter estimation result for one test instance with a signal-to-noise ratio (SNR) of 501.2 is shown in Fig.~\ref{fig:corner}, where the brown contour plot represents the estimation result of the CVAE model, and the blue contour plot represents the result obtained from the Nested Sampling.

\subsection{Reducing Prior Range}
\label{sec:nsresults}

In the Bayesian framework, where the prior $\pi(\theta)$ is non-zero but the likelihood $\mathcal{L}(d|\theta)$ is zero, the posterior $p(\theta|d)$ will be zero. If the prior is also set to zero in such regions, the posterior will be unaffected. Therefore, it is reasonable to assign zero prior probability to the region where the prior is non-zero but the likelihood is very small. This strategy can effectively reduce the scope of the prior without substantially affecting the overall posterior distribution and is largely equivalent to bypassing the generation of low-weight sampling points in the early iterations of Nested Sampling.

The trained CVAE model estimates the parameters of a single instance in approximately one second, while the Nested Sampling method, using parallel computation across 16 CPU cores with the \texttt{PyMultiNest} sampler in the Bilby framework~\cite{ashton2019bilby, romero2020bayesian}, requires an average of 20 hours for one test instance. Although the trained model has relatively weaker parameter constraint capabilities, its significant advantage in sampling speed over Nested Sampling makes it suitable for preprocessing the prior of Nested Sampling. By narrowing the prior range, the initial sampling space of the standard Bayesian sampling is reduced, thereby accelerating the sampling process. 

For each parameter of GW, we take the intersection between the central $99.73\%$ confidence interval of the CVAE results, defined as the range from the 0.135th percentile to the 99.865th percentile, and the initial full prior range specified in Table~\ref{tab:table1}, as the narrowed prior range. The reason for introducing the constraint of the initial full prior is that in some cases, the results generated by the CVAE may exceed the original full prior range. For example, in Fig.~\ref{fig:corner}, the upper bound of $d_L$ and the upper and lower bounds of $\phi$ and $\psi$ in CVAE results exceed the initial full prior range. However, these exceeded regions will not be considered even under standard Bayesian sampling with the full prior. Therefore, removing these regions does not require any information beyond what the standard Bayesian sampling includes. The gray region in Fig.~\ref{fig:corner} shows the narrowed prior range obtained by the above intersection.

The Nested Sampling result after prior narrowing for the GW instance in Fig.~\ref{fig:corner} is nearly identical in accuracy and precision to that obtained under the full prior, as shown in Fig.~\ref{fig:corner_2} of Appendix~\ref{app:narrowed_prior}. This indicates that the CVAE-reduced prior does not significantly impact the posterior distribution of the instance. With CVAE preprocessing, the Nested Sampling runtime for the instance is reduced from 47.5 hours to 3.9 hours under parallel computation, corresponding to a reduction of a factor of $\sim$12. Across the 50 instances in the test set, the runtime is reduced by a factor of $\sim$6 on average.

\begin{figure}
\includegraphics[width=8.6cm]{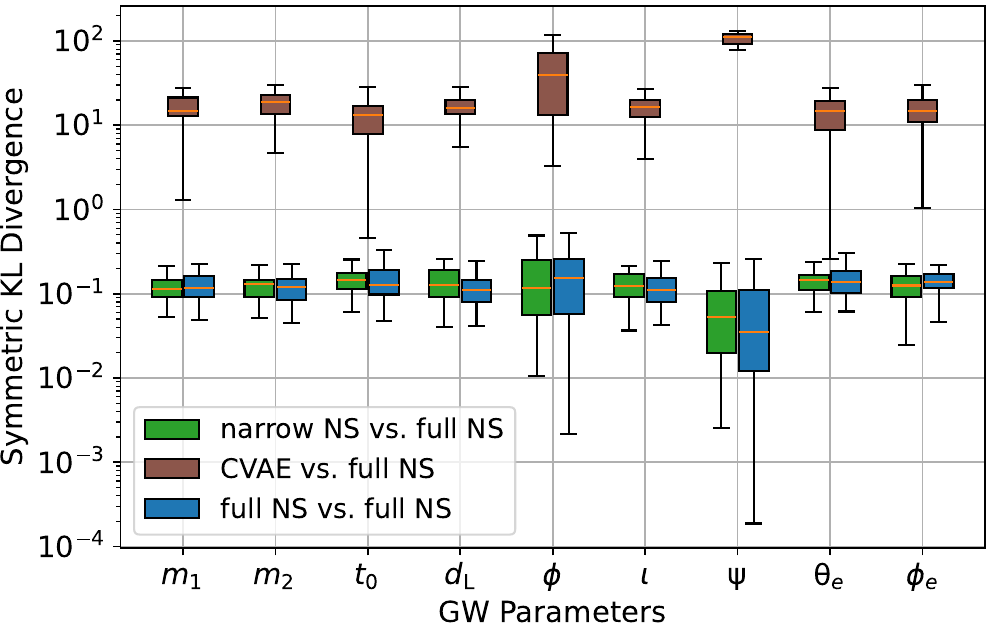}
\captionsetup{justification=raggedright,singlelinecheck=false}
\caption{\label{fig:kl} Boxplot illustrating the $D_{s\mathrm{KL}}$ distribution of 50 GW instances across different methods. The box represents the interquartile range, which spans from the 25th percentile to the 75th percentile. The orange line inside the box represents the median. The whiskers extend to the most extreme data points within 1.5 times the interquartile range from the box. We present a boxplot (green) showing the $D_{s\mathrm{KL}}$ between Nested Sampling runs with CVAE-reduced prior (narrow NS) and with complete prior (full NS). This is compared with the boxplots of \texttt{"CVAE vs.~full NS"} (brown) and \texttt{"full NS vs.~full NS"} from two repeated runs (blue).}
\end{figure}

We use the symmetric KL divergence $D_{s\mathrm{KL}}$ to quantify the discrepancy between posterior distributions obtained from different sampling processes. The definition of $D_{s\mathrm{KL}}$ and the corresponding sample-based estimation method are provided in Appendix~\ref{app:skl}. The $D_{s\mathrm{KL}}$ between two Nested Sampling results before and after narrowing the prior across 50 instances for 9 source parameters is shown in Fig.~\ref{fig:kl}, labeled as \texttt{"narrow NS vs.~full NS"}. 
A $D_{s\mathrm{KL}}$ of 0 indicates perfect similarity between two distributions. However, due to sampling uncertainty, the $D_{s\mathrm{KL}}$ between two separate Nested Sampling runs of the same instance is typically slightly greater than 0. To quantify this uncertainty, we perform two independent runs with full prior, and calculate the corresponding $D_{s\mathrm{KL}}$ across 50 instances, labeled as \texttt{"full NS vs.~full NS"}. Furthermore, we compare the posterior estimated by the CVAE model with the Nested Sampling posterior obtained with the full prior. This comparison is labeled as \texttt{"CVAE vs.~full NS"}.

By comparing the $D_{s\mathrm{KL}}$ of \texttt{"CVAE vs.~full NS"} and \texttt{"full NS vs.~full NS"}, we find that the $D_{s\mathrm{KL}}$ of \texttt{"CVAE vs.~full NS"} is larger than that of \texttt{"full NS vs.~full NS"}. This indicates that the estimation accuracy of the CVAE model is significantly inferior to that of the standard sampling method. Additionally, by comparing the $D_{s\mathrm{KL}}$ of \texttt{"narrow NS vs.~full NS"} and \texttt{"full NS vs.~full NS"} we find that the two have a similar magnitude, suggesting that narrowing the prior has minimal impact on the Nested Sampling results. These results highlight the effectiveness of using deep learning as a preprocessing step to accelerate parameter estimation.

\section{conclusions}
\label{sec:conclu}
In this study, we employ the CVAE model to estimate the posterior of source parameters for unknown time-domain MBHB signals. The variational nature of CVAE enables the generation of complex parameter distributions. The model is trained for $2$ days on an A800 GPU using $6\times 10^5$ training instances and $1.2\times 10^4$ validation instances, utilizing approximately 68~GB of GPU memory during training. Following training, the model is evaluated on the test set comprising 50 instances to obtain their posterior distributions. Standard Bayesian sampling for GW parameters is computationally intensive, typically requiring significant time and resources. In contrast, the trained CVAE model demonstrates superior efficiency, with an average sampling time of approximately one second per instance, compared to the standard posterior sampling method, which takes 20 hours on average when executed through parallel processing on a CPU with 16 cores.

However, compared to the standard posterior sampling method, the CVAE results exhibit a certain degree of light-tailed behavior, with the distribution widths spanning several orders of magnitude relative to the standard Bayesian results. Given that CVAE processes MBHB signals in approximately one second, we propose that for scenarios requiring higher precision in parameter estimation, CVAE can be employed to preprocess GW signals to narrow down the prior range of the standard Bayesian sampling method. By selecting the parameter interval as the intersection between the central $99.73\%$ confidence interval of CVAE and the full prior range, a more restricted prior range can be obtained to accelerate the standard posterior sampling process.

In our experiments across 50 test instances, the runtime of standard posterior sampling with the narrowed prior is on average reduced by a factor of $\sim$6 compared to that with the full prior. Furthermore, by evaluating the symmetric KL divergence between the sampling results before and after narrowing down the prior range, we show that the level of agreement is comparable to that observed in the two independent full Nested Sampling runs.

This method can also be applied to other gravitational wave sources with high signal-to-noise ratios, where differences remain in the precision of parameter estimation between the deep learning method and the standard Bayesian sampling method.

Added-in-proof. While finalizing the draft, we become aware of a recently submitted related study~\cite{nerin2024parameter} that also employs CVAE-generated priors for Bayesian sampling. Their work focuses on lensed gravitational wave signals from stellar-mass black hole binaries, specifically targeting two lensing-related parameters.

\begin{acknowledgments}

The author Hui Sun would like to acknowledge the helpful discussions with Junshuai Wang on parallel computing and GPU-based model training. This research is funded by the Strategic Priority Research Program of the Chinese Academy of Sciences under Grant No. XDA15021100, as well as the Fundamental Research Funds for the Central Universities. HW is partially supported by the National Key Research and Development Program of China (Grant No. 2021YFC2203004) and National Science Foundation of China (NSFC) under Grant No. (12405076).

\end{acknowledgments}

\appendix

\section{Effect of Prior Narrowing on the Nested Sampling Posterior}
\label{app:narrowed_prior}

\begin{figure*}[ht]
\includegraphics[width=17cm]{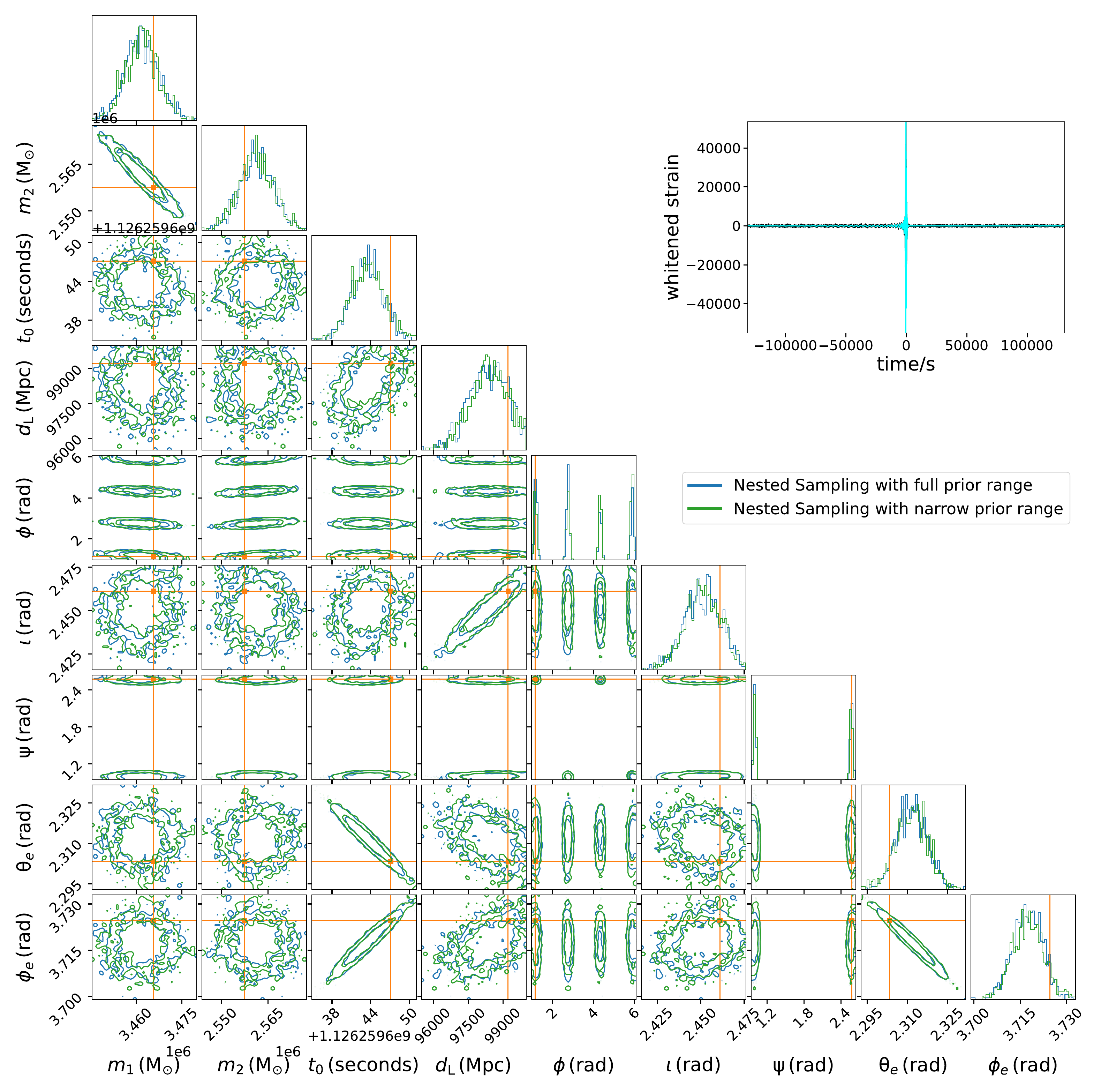}
\caption{\label{fig:corner_2} Distributions of source parameters obtained using the standard Bayesian sampling method under the initial full prior (blue) and the narrowed prior (green). The result contours of each method correspond to the $68\%$ and $95\%$ confidence levels. The orange solid lines are the true parameter values of the injected instance.}
\end{figure*}

Figure~\ref{fig:corner_2} shows the Nested Sampling results for the GW instance in Fig.~\ref{fig:corner}, comparing the posteriors obtained with and without CVAE-based prior narrowing. The two distributions exhibit nearly identical accuracy and precision. 
 
\section{Symmetric KL Divergence}
\label{app:skl}
 
We introduce the symmetric KL divergence to describe the discrepancy between distributions $A$ and $B$ : 
 
\begin{equation}
D_{s\mathrm{KL}}(A\parallel B) \equiv D_{\mathrm{KL}}(A\parallel B)+D_{\mathrm{KL}}(B\parallel A).
\end{equation}
Assume we have two sets of samples: $N$ samples $x_i$ and $N^{\prime}$ samples $x^{\prime}_j$, where $x_i \sim A(x)$ and $x^{\prime}_j \sim B(x)$. The KL divergence between the distributions $A$ and $B$, and between $B$ and $A$, can be approximated as
 
\begin{align}
D_{KL}\left(A(x) \| B(x)\right) & \approx \frac{1}{N} \sum_{i=1}^N \log \frac{\hat{A}(x_i)}{\hat{B}(x_i)},  \\
D_{KL}\left(B(x) \| A(x)\right) & \approx \frac{1}{N^{\prime}} \sum_{j=1}^{N^{\prime}} \log \frac{\hat{B}(x^{\prime}_j)}{\hat{A}(x^{\prime}_j)}.
\end{align}
Where $\hat{A}(x)$ and $\hat{B}(x)$ are the distributions estimated based on the samples $x_i$ and $x^{\prime}_j$, respectively. In this work, we use the Gaussian kernel density estimation method \cite{parzen1962estimation} to fit the samples sets $x_i$ and $x^{\prime}_j$, and obtain $\hat{A}(x)$ and $\hat{B}(x)$.

\nocite{*}

\bibliography{main}

\end{document}